\documentclass[sigconf,
balance,
]{acmart}
\usepackage{amsfonts}
\usepackage{bbm}
\usepackage{booktabs}
\usepackage{makecell}
\usepackage{tikz}
\usepackage{adjustbox}
\usepackage{enumitem}

\usepackage{fancyhdr}
\newif\ifshowcomment
\showcommenttrue
\newcommand{\ed}[1]{\textcolor{black}{#1}}

\def\systemnameRaw{\textsc{EnergAt}}
\def\systemname{\systemnameRaw}

\newcommand{\tocite}[1]{[{\textcolor{red}{?}}]}
\newcommand{\toref}[1]{\textcolor{red}{N}}
\ifshowcomment
\newcommand{\TODO}[1]{\textcolor{red}{{[\small\textsf{{TODO: #1}}}]}}
\newcommand{\NOTE}[1]{\textcolor{orange}{{[\small\textsf{{NOTE: #1}}}]}}
\else
\newcommand{\TODO}[1]{}
\newcommand{\NOTE}[1]{}
\fi

\newcommand{\pr}[0]{\mathbb{P}}
\newcommand{\E}[0]{\mathbb{E}}
\newcommand{\indicator}[0]{\mathbbm{1}}

\newcommand{\circleblack}[1]{%
 \begin{tikzpicture}[baseline=(char.base)]
   \node[draw,circle,inner sep=0.5pt, fill=black, text=white] (char){\small #1};
 \end{tikzpicture}%
 }

\AtBeginDocument{%
  }

\setcopyright{acmcopyright}
\copyrightyear{2023} 
\acmYear{2023} 
\setcopyright{acmlicensed}\acmConference[HotCarbon '23]{2nd Workshop on Sustainable Computer Systems}{July 9, 2023}{Boston, MA, USA}
\acmBooktitle{2nd Workshop on Sustainable Computer Systems (HotCarbon '23), July 9, 2023, Boston, MA, USA}
\acmPrice{15.00}
\acmDOI{10.1145/3604930.3605716}
\acmISBN{979-8-4007-0242-6/23/07}

\begin{document}

\title{\systemname{}: Fine-Grained Energy Attribution for Multi-Tenancy}


\author{Hongyu Hè}
\authornote{Corresponding author <\texttt{hongyu.he@inf.ethz.ch}>.}
% \email{honghe@inf.ethz.ch}
\affiliation{%
  \institution{ETH Zurich}
  % \city{Z\"urich}
  \country{Switzerland}
}

\author{Michal Friedman}
% \email{michal.friedman@inf.ethz.ch}
\affiliation{%
  \institution{ETH Zurich}
  % \city{Z\"urich}
  \country{Switzerland}
}

\author{Theodoros Rekatsinas}
% \email{}
\affiliation{%
  \institution{Apple Inc.}
  % \city{Z\"urich}
  \country{Switzerland}
}







%-------------------------------------------------------------------------------
\begin{abstract}
%-------------------------------------------------------------------------------

In the post-Moore's Law era, relying solely on hardware advancements for automatic performance gains is no longer feasible without increased energy consumption, due to the end of Dennard scaling. 
Consequently, computing accounts for an increasing amount of global energy usage, contradicting the objective of sustainable computing.
The lack of hardware support and the absence of a standardized, software-centric method for the precise tracing of energy provenance exacerbates the issue. 
Aiming to overcome this challenge, we argue that fine-grained software energy attribution is attainable, even with limited hardware support.
To support our position, we present a thread-level, NUMA-aware energy attribution method for CPU and DRAM in multi-tenant environments. 
The evaluation of our prototype implementation, \systemname{}, demonstrates the validity, effectiveness, and robustness of our theoretical model, even in the presence of the noisy-neighbor effect.
We envisage a sustainable cloud environment and emphasize the importance of collective efforts to improve software energy efficiency.

% The end of Dennard scaling has turned the spotlight on software energy efficiency.
% In the post-Moore's Law era, programs can no longer rely on automatic performance gains from hardware advancements without burning more energy, which has led to growing attention on software energy efficiency.
% As a result, computing accounts for an increasingly significant amount of global energy usage, which is antithetical to the goal of sustainable computing.
% Unfortunately, neither does existing hardware support fine-grained energy attribution, nor is there a standard software-centric method for tracing detailed energy provenance.
% To address this challenge, we argue that fine-grained software energy attribution is feasible even with limited hardware support.
% To support our stance, we propose and present promising results of a thread-level, NUMA-aware energy attribution method for CPU and DRAM in multi-tenant environments.
% Through our prototype implementation named \systemname{}, we demonstrate the validity, effectiveness, and robustness of our theoretical model, even in the presence of the noisy-neighbor effect.
% We conclude by laying out our vision of a sustainable cloud environment and emphasizing the need for collective efforts in improving software energy efficiency.
\end{abstract}
\maketitle

\newcommand{\sd}[1]{{\color{red}#1}} %comments from Shail

% \balance 

%-------------------------------------------------------------------------------
\section{Introduction}
%-------------------------------------------------------------------------------

    % \begin{figure}[t]
    %     \centering
    %     \begin{adjustbox}{width=1.15\linewidth,center=0pt}
    %       \includegraphics[width=\linewidth]{figures/tools.pdf}
    %     \end{adjustbox}
    %     \caption{Drastically different results from various tools on four microbenchmarks (Table~\ref{tab:benchmarks}). We highlight three observations: (1) existing methods can exhibit more than 46.3\% overestimation and 93.3\% underestimation of the \textit{attributed} energy compared to the \textit{total} values, (2) the total energy consumption measured by our fine-grained energy attribution model \systemname{} matches that of the reference value, and (3) the dotted line illustrates that \systemname{} is robust to the noisy-neighbor effect.}
    %     \label{fig:tools}
    % \end{figure}

In the post-Moore era, accelerating application by hardware advancements typically requires computing systems to burn more energy. 
This leads to increasing amounts of global energy consumption~\cite{jones2018stopdc,masanet2020recalibrating,shehabi2016usenergyreport,DBLP:journals/corr/odcsim}, which impairs the sustainability of computing operations. Unlike hardware, software design and optimization often neglect their impact on energy efficiency and carbon footprints. This phenomenon is primarily due to challenges in \textit{software energy attribution}. Energy attribution of software aims to determine the share of energy consumed by the target application and its subtasks (\textit{energy provenance}), excluding the fraction used by other collocated applications. 
Task-level energy attribution in multi-tenant environments could not only facilitate energy-aware decision-making in the cloud but also help developers gain first-hand insights into the energy efficiency and carbon footprint of their applications.
% \sd{TODO: Write why software-level or fine-grained energy attribution is important. We can reuse material from your text below on what we are missing for coarse-grain attribution.} 

However, such fine-grained attribution is particularly hard due to the lack of support from both hardware and software.
First, energy-related statistics measured by hardware are typically \textit{coarse-grained} (at the device/socket-level) and does not support runtime or software-level information (fine-grained) for energy attribution. For instance, the energy consumed by a program running on a CPU core for a period of time cannot be directly obtained from hardware, since CPU power is measured at the socket level~\cite{DBLP:journals/micro/interlraplvalidation}.
Second, the problem of software attribution is exacerbated by the increasing heterogeneity and multi-tenancy in the cloud. For example, together with CPUs, GPUs, and storage, specialized accelerators such as TPUs and FPGAs are increasingly shared among many tenants in the cloud~\cite{DBLP:conf/osdi/coyote,DBLP:conf/asplos/optimus,DBLP:conf/micro/cloudscale,fracgpu,DBLP:conf/hpdc/sharegpu,jouppi2023tpuv4,hennessy2019goldenca}. 
Moreover, the push for compute-storage disaggregation~\cite{bindschaedler2020hailstorm,angel2020disaggregation} complicates the problem of accurately attributing energy at the application/subtask level further.
Therefore, hardware support alone cannot effectively solve the problem of fine-grained energy attribution, and software solutions are needed.

\vspace{-6pt}
\paragraph{\textbf{Gap.}}
In recent years, several tools have been developed to measure software energy consumption (e.g., \cite{DBLP:journals/corr/codecarbon,DBLP:journals/corr/carbontracker,DBLP:journals/corr/impacttracker,DBLP:journals/corr/pmt,scaph,DBLP:conf/asplos/powercontainer}).
These tools have primarily focused on usability, accessibility, and the interpretability of their outputs, but they do not aim for fine-grained energy attribution.
Specifically, they either assume that the measured application is not collocated with other tasks, treating the total energy consumption of the host machine as that of the target application, or they use coarse-grained energy attribution models at the process level~\cite{colmant2015process,DBLP:conf/asplos/powercontainer} and do not consider NUMA effects in case of multiple CPU sockets (\S\ref{sec:sw-acct}).
Moreover, these tools do not update the traced subtasks of the target application reactively, which is problematic since processes and threads can be created and deleted at runtime.
Furthermore, these tools do not separate their own energy cost from the measurements of target. 
We find that the lack of such accounting in fine-grained, software-centric attribution methodology~\cite{anderson2022treehouse} leads to more than 46.3\% overestimation and 93.3\% underestimation (Fig.~\ref{fig:tools}), which could be detrimental for sustainable runtime operations. 

\vspace{-3pt}
\paragraph{\textbf{Our position.}}
We argue that developing fine-grained energy attribution models is feasible even with coarse-grained hardware support. 
These models could serve as the foundation for building sustainable cloud environments.
In order to achieve this objective, precise and validated accounting of software energy provenance is required in multi-tenant environments. 
To support our stance with concrete exposition, we make the following contributions: 
\begin{enumerate} [label={{(\arabic*)}},leftmargin=10pt,labelsep=3pt] % [label={{(\arabic*)}},align=left,itemindent=0pt,listparindent=0pt,leftmargin=25pt,labelsep=0pt,topsep=0pt,partopsep=0pt,parsep=0pt,itemsep=0pt]
    \item We propose a thread-level, NUMA-aware method for CPU and DRAM energy attribution in a multi-tenant environment (\S\ref{sec:method}).
    \item To evaluate our theoretical model, we provide a prototype implementation (\S\ref{sec:impl}) and present preliminary results (\S\ref{sec:eval}) that demonstrate the attribution model's validity, effectiveness, and robustness to the noisy-neighbor effect in multi-tenant environments.
    \item Building upon the fine-grained tracing of software energy provenance, we envisage a cloud environment in which {decentralized}, software-centric energy attribution supports a \textit{logically} centralized runtime control to conduct energy-aware operations and provide feedback to users about their applications' energy efficiency (\S\ref{sec:future-work}). We also present insights into promising opportunities and prominent challenges thereof.
\end{enumerate}
\enlargethispage{\baselineskip}

\section{Coarse-Grained Energy Attribution} \label{sec:sw-acct}

% \vspace{-3pt}
\paragraph{\textbf{Coarse hardware support.}}
Commodity hardware mostly supports coarse-grained power estimation.
For instance, as of the Sandy Bridge generation, Intel processors come with a built-in power meter, the Running Average Power Limit (RAPL)~\cite{DBLP:journals/micro/interlraplvalidation}.
It provides an interface to the accumulated energy consumption of various components, e.g., CPU package and DRAM. 
Although its energy reporting is mainly based on kernel events~\cite{DBLP:journals/micro/interlraplvalidation}, RAPL has been proven to be sufficiently accurate and can capture millisecond-level energy events~\cite{DBLP:journals/sigmetrics/shortcodepath,DBLP:journals/tompecs/raplinaction,DBLP:journals/micro/interlraplvalidation}.
GPUs also support power reporting, e.g., NVIDIA's NVML~\cite{nvml} and ROCm SMI~\cite{rocm} from AMD.
For FPGAs, Xilinx-AMD provides dedicated tools for power estimation~\cite{xpe}. 
Unfortunately, none of these devices inherently supports fine-grained energy measurement and attribution.
For example, energy measurements from RAPL are reported at the socket level.
Similarly, although fractional sharing of GPU~\cite{fracgpu} and FPGA~\cite{DBLP:conf/asplos/optimus} is now possible, their energy reporting is still at the device/user level.

\vspace{-3pt}
\paragraph{\textbf{Coarse-grained software energy accounting.}} \label{par:tools}
Several recent tools have been developed to measure software energy consumption.
Many of them focus on machine learning (ML) workloads, which are particularly energy-intensive~\cite{DBLP:conf/mlsys/metaai,DBLP:journals/corr/impacttracker,DBLP:journals/corr/googleads}.
To name a few, CodeCarbon~\cite{DBLP:journals/corr/codecarbon} reports the energy usage of a program, measuring the consumption of CPU, DRAM, and GPU. 
It features a user-friendly API, a UI and exports interpretable results.
Carbontracker~\cite{DBLP:journals/corr/carbontracker} offers similar measurements and predicts the energy consumption of ML training based on a few epochs.
Specifically built for ML applications, Experiment Impact Tracker (Impact Tracker)~\cite{DBLP:journals/corr/impacttracker} collects energy measurements for both CPU and GPU, and allows users to generate environmental impact statements for their experiments.
Another popular tool is Scaphandre~\cite{scaph} which has integrated support for power measurements in Kubernetes.
% For virtualized environments where applications are treated as black boxes (e.g., containers and VMs), predictive models have been proposed to estimate their energy usage (e.g., \cite{fieni2020smartwatts,bohra2010vmeter,krishnan2011vmmetering,dhiman2010predictvirtual}).
% \vspace{-3pt}
% \paragraph{Coarse-grained energy accounting.}

Unfortunately, existing tools only employ coarse-grained energy attribution models as summarized below.
Consider a target application $\mathcal{A}$ that potentially consists of multiple tasks $a$ running on a set of devices $D$. 
Then, \textit{coarse-grained} energy attribution models can be formulated as:
\begin{align}
    E_\text{total}^D &\gets \text{Sample power meter of }{D} \text{ every } T_\text{sample} \label{eq:ex-total} \\
    % C^{\text{CPU}} &= \left( \sum_{a \in \mathcal{A}} T_a^{\text{CPU}} \right) / T_\text{total}^{\text{CPU}} \label{eq:ex-CPUcredit} \\
    % C^{\text{DRAM}} &= \left( \sum_{a \in \mathcal{A}} M_a \right) / M_\text{total} \label{eq:ex-memcredit} \\
    E_\mathcal{A}^D &= E_\text{total}^D \cdot \left[ \left( \sum\nolimits_{a \in \mathcal{A}} U_a^{D} \right) \big/ U_\text{total}^{D} \right], \label{eq:coarse-model}
    % E_\mathcal{A}^\text{DRAM} &= E_\text{total}^\text{DRAM} \cdot \left[ \left( \sum\nolimits_{a \in \mathcal{A}} M_a \right) / M_\text{total} \right], \label{eq:ex-mem} 
\end{align}
where $E_\text{total}^D$ and $E_\mathcal{A}^D$ are the total energy consumed by the server machine and application $\mathcal{A}$ on $D$, respectively. 
Every $T_\text{sample}$ time, the corresponding energy meter of the device is sampled to obtain the accumulated energy for this period.
$U_a^D$ and $U_\text{total}^D$ are the resource usage of the application task $a$ and that of all tasks present on the server machine.
We argue that this energy attribution method is insufficient in precisely capturing applications' power dynamics.
Specifically, it neither takes into account NUMA effects~\cite{DBLP:conf/usenix/numaimbalance,lepers2015numaasym,zhang2013numaisolate,dashti2013numatraffic} in the presence of multiple sockets nor does it distinguish threads from processes when tracing energy provenance~(\S\ref{subsec:factors}).
This method is also prone to the noisy-neighbor effect~\cite{DBLP:conf/iiswc/ibench,DBLP:conf/asplos/bolt} in a multi-tenant environment, where the energy attribution of an application is interfered by collocated tasks on the same host.
\enlargethispage{\baselineskip}
\section{Thread-Level and NUMA-Aware Energy Attribution} \label{sec:method}

Although present-day hardware only provides coarse-grained energy measurement capabilities, we believe that there is still immense potential to achieve fine-grained energy attribution with a software-based approach~\cite{anderson2022treehouse}.
In this work, we demonstrate fine-grained energy attribution of CPU and DRAM with coarse-grained measurements from Intel RAPL meter.
CPU and DRAM are primary consumers of software energy. 
Even for GPU-dependent ML applications, they together still account for more than 30\% of total energy use~\cite{DBLP:conf/mlsys/metaai,DBLP:journals/corr/carbontracker}.
% That said, our methodology offers insights into the fine-grained energy attribution with other computing/memory devices as well.
% In what follows, we present a thread-level, NUMA-aware method that traces software energy provenance robust to multi-tenancy with only the current hardware support.

\subsection{Relevant Factors in Energy Attribution} \label{subsec:factors}

% \vspace{-3pt}
\paragraph{\textbf{Multiple sockets.}}

Server-class machines generally have $\ge2$ CPU sockets, e.g., for higher memory capacity and fault tolerance.
Unfortunately, when measuring and attributing energy, prior work~(\S\ref{par:tools}) does not take into account NUMA effects, and the application resources are aggregated over all sockets (Eq.~\ref{eq:coarse-model}).
This approach can be problematic for accurate attribution due to the potential imbalance of resource allocation in NUMA architectures~\cite{DBLP:conf/usenix/numaimbalance}.
For example, a dual-socket machine has only one user task running, whose CPU times are 30 s and 180 s on each of the two sockets.
If the \textit{total} CPU times and the measured energy consumption of the two sockets are (100 s, 30 J) and (200 s, 50 J) respectively, then the CPU energy consumption attributed to the task should be $\left(30/100 \times 30 + 180/200 \times 50\right)$ J, instead of $[(30+180)/(100+200) \times (30+50)]$ J.
\ed{
Note that, apart from CPU time, another crucial factor making a difference here is CPU utilization.
Specifically, utilization typically varies among sockets at a certain point in time, and power scales non-linearly with it~\cite{DBLP:conf/www/nonlinear,DBLP:journals/computer/newenergyprop,DBLP:journals/taco/paretogovernors}.
Consequently, the same amount of CPU time would result in different power dynamics at different utilization levels.
Therefore, {relying solely on CPU time as an aggregated proxy is insufficient for estimating energy consumption across multiple sockets}.}
The same principle also applies when accounting for memory's energy consumption. 
\ed{In practice, however, energy variability caused by NUMA memory allocation plays a less significant role, compared to CPU.}
%, e.g., at a certain point in time, an application allocates 10\% of its memory on one socket and the rest 90\% on another.
% Therefore, software energy attribution should take into account NUMA effects.

\vspace{-3pt}
\paragraph{\textbf{Threads vs. processes.}}
High parallelism is prevalent in modern applications. 
\ed{Apart from traditional high-performance computing workloads}, large numbers of parallel tasks in ML pipelines (e.g., feature stores~\cite{DBLP:journals/pvldb/featurestore,Michelangelo,ralf}) can amount to 1/3$^{rd}$ of the total energy consumption, exceeding the amount of energy used by model training of large-scale jobs~\cite{DBLP:conf/mlsys/metaai}.
% \ed{Moreover, threads are frequently used in these applications as they can provide additional concurrency in virtualized environments~\cite{kotni2021faastlane}.}
To precisely attribute CPU energy for a parallel application, one needs to obtain the CPU time for each of its tasks (processes \textit{and} kernel threads) per socket.
For example, a parallel application spawns several threads and processes, each of which could have different runtime statistics on different sockets, depending on the placement decisions made by the scheduler.

Furthermore, when querying resource statistics, tools such as \texttt{ps} and \texttt{top} either (1) use the total resource usage of the process group (PG)\footnote{Resource group created by the parent process.} as that of a single task therein (process/thread), or (2) separate the statistics for each task, given different flags.
Unfortunately, existing libraries do not always handle the two cases properly.
For example, many energy tools rely on the library \texttt{psutil}~\cite{psutil}, which reports total resource usage of the PG when asked for that of a thread (case (2)).
In turn, when tracing energy provenance at the process level, case (1) can result in imprecise measurements (since the resource usages of threads and processes cannot be distinguished for individual accounting), and case (2) can lead to underestimation (as only the resource usage of processes is accounted and that of threads is ignored).
Thus, the fact that $a$ in Eq.~\ref{eq:coarse-model} purely represents processes and ingores threads is problematic for fine-grained attribution.
Note that, for resources shared between the parent process and its threads (e.g., stack memory), making such a distinction is unnecessary.
However, energy tracking tools themselves should \textit{not} be created as threads of the application it measures. 
Otherwise, the resources used by the energy tool would entangle with its target application, which is also a pitfall in existing methods.

\vspace{-3pt}
\paragraph{\textbf{Noisy-neighbor effects.}}
Nowadays, applications are typically deployed in the cloud, where they are colocated with other tasks, sharing resources on the same host.
Multi-tenancy creates ``noisy-neighbor'' effects~\cite{DBLP:conf/iiswc/ibench,DBLP:conf/asplos/bolt}, by which the performance of an application is interfered by its ``neighbors''.
In the presence of such interference, only the energy consumed by the target application should be accounted.
% We argue that such interference should be dealt with care when tracing software energy provenance.
% In other words, the attribution model should only attribute energy consumed by the target application, excluding that of other collocated neighbors.

\subsection{Fine-Grained Attribution Model}
Taking into account the aforementioned factors (\S\ref{subsec:factors}), we propose a thread-level energy attribution model that is NUMA-aware and robust to the noisy-neighbor effect.

\vspace{-3pt}
\paragraph{\textbf{Static power.}}
The first step in our model is to measure the static power of the host on which the target application runs.
\ed{The static power is assumed to be independent of the load and should not be confused with the idle power, which is consumed by the server machine in various sleep states~\cite{Heinrich2017PredictingEnergyMPI,DBLP:journals/corr/odcsim,cstates}. 
This value can either be obtained from the manufacturer's datasheet or more practically, via a sampling procedure.}
For each socket $s \in S$ and a sampling period $T_\text{static}$, the average static power $\left(P_\text{static}\right)^s$ is given by:
\begin{align}
    \left(P^D_\text{static}\right)^s &= \left(\text{Sample energy value of } D \text{ for }T_\text{static}\right) / T_\text{static}. \label{eq:static}
\end{align}
Then, at runtime, the total energy used by the host $\left(E_\text{total}^D\right)^s$ can be obtained using Eq.~\ref{eq:ex-total} for each $s$, and the static energy consumption $\left(E^D_\text{static}\right)^s$ can be obtained periodically:
\begin{align}
    \left(E^D_\text{static}\right)^s &= \left(P^D_\text{static}\right)^s \cdot T_\text{sample}. \label{eq:static}
    % \left(E_\text{total}^D\right)^s &= \text{Sample}_D(T_\text{sample}). \label{eq:sample}
\end{align}

\vspace{-3pt}
\paragraph{\textbf{Attributing CPU energy with thread-level metrics.}}
To attribute CPU energy, we first obtain the power offset $\left(E_\Delta^\text{CPU}\right)^s$ by subtracting the static power from the total:
\begin{align}
    \left(E_\Delta^\text{CPU}\right)^s &= \left(E_\text{total}^\text{CPU}\right)^s - \left(E_\text{static}^\text{CPU}\right)^s. \label{eq:delta-cpu}
\end{align}
Having obtained the host-wide energy statistics, we now quantify the resource usage of $\mathcal{A}$ in \textit{thread} granularity.
Specifically, we estimate the \textit{CPU residence rate} for every process \textit{and} thread $a \in \mathcal{A}$ on each socket $s$, 
i.e., the fraction of time task $a$ was scheduled on $s$:
\begin{align}
    \pr^\text{CPU}(s \mid a) \approx \left(\int_{t=t'}^{t'+T_\text{sample}} \indicator_{\{a\ \mathrm{on}\ s\}} dt \right) \bigg/ T_\text{sample}, \label{eq:residence-cpu}
\end{align}
where $dt$ in practice is the discretized time steps for sampling kernel scheduling decisions.
With Eq.~\ref{eq:residence-cpu}, we approximate the CPU time of $\mathcal{A}$ on $s$ with an expectation conditioned on the kernel scheduling decisions:
\begin{align}
    \left( T_\mathcal{A}^\text{CPU} \right)^s = \E\left[T_\mathcal{A}^\text{CPU} \mid s\right] \approx \sum_{a \in \mathcal{A}} \pr^\text{CPU}(s \mid a) \cdot T_a^\text{CPU}, \label{eq:cpu-time}
\end{align}
where $T_a^\text{CPU}$ is the total CPU time of $a$ on all $s \in S$.

Furthermore, to combat the noisy-neighbor effect, we propose the concept of \textit{energy credit} denoted $C_\mathcal{A}^D$, that is, how much a fraction of the energy consumption of $D$ should be attributed to $\mathcal{A}$.
Specifically, we employ the proportion of $\mathcal{A}$'s CPU time over that of all tasks on the server as a proxy for the CPU energy credit on $s$:
\begin{align}
        (T_\text{total})^s &\gets \text{Total CPU time (kernel + user) of }s \label{eq:total-cputime} \\
    \left(C^\text{CPU}_\mathcal{A}\right)^s &= \left[ \left( T_\mathcal{A}^\text{CPU} \right)^s \big/ \left( T_\text{total}^\text{CPU} \right)^s \right]^\gamma, \label{eq:cpu-credit}
\end{align}
where $(T_\text{total})^s$ is the server-wide CPU time per socket and $0\geq \gamma \leq 1$ is a scaling factor that takes into account machine-specific non-linearity, since the energy consumption of CPU does not scale linearly with the utilization~\cite{DBLP:conf/www/nonlinear,DBLP:journals/computer/newenergyprop,DBLP:journals/taco/paretogovernors}.
Specifically, the trend flattens gradually as utilization gets higher.
Using the CPU energy credit, we compute the energy consumption of $\mathcal{A}$ by aggregating values from all sockets:
\begin{align}
    E_\mathcal{A}^\text{CPU} = \sum_{s \in S} \left(E_\Delta^\text{CPU}\right)^s \cdot \left(C^\text{CPU}_\mathcal{A}\right)^s + \left(E_\text{static}^\text{CPU}\right)^s. \label{eq:cpu}
\end{align}

\vspace{-3pt}
\paragraph{\textbf{Attributing DRAM energy with NUMA memory statistics.}}
The energy attribution for DRAM is similar, except that we no longer consider threads individually, since they share memory under the same PG.
However, the imbalance in memory allocation of NUMA architectures still needs to be dealt with care.
Firstly, we obtain the server-wide memory usage per socket $(M_\text{total})^s$ and the memory offset $\left(E_\Delta^\text{DRAM}\right)^s$: %, akin to Eq.~\ref{eq:total-cputime}, \ref{eq:delta-cpu}:
\begin{align}
    (M_\text{total})^s &\gets \text{Total available NUMA memory on } s \label{eq:total-mem} \\
    \left(E_\Delta^\text{DRAM}\right)^s &= \left(E_\text{total}^\text{DRAM}\right)^s - \left(E_\text{static}^\text{DRAM}\right)^s. \label{eq:delta-mem}
\end{align}
Next, we measure \textit{memory residence rate} for each \textit{process} $a \in \mathcal{A}$, i.e., the fraction of \textit{private} NUMA memories allocated on $s$, excluding shared memories (whose ownership is hard to reason about):
\begin{equation}
    \pr^\text{DRAM}(s \mid a) \approx \E \left[ \left\{  \left(M_a^{\Delta t}\right)^s \bigg/ \left(M_\text{total}^{\Delta t}\right)^s \right\}^{T_\text{sample}} \right], \label{eq:residence-mem}
\end{equation}
where $\{\}$ encloses a collection of memory samples on $s$ over a sampling period $T_\text{sample}$ with discretized steps $\Delta t$.
Then, the total NUMA memory of $\mathcal{A}$ on $s$ can be approximated by:
\begin{equation}
    (M_\mathcal{A})^s =  \E\left[M_\mathcal{A} \mid s\right] \approx \sum_{a \in \mathcal{A}} \pr^\text{DRAM}(s \mid a) \cdot (M_a)^s. \label{eq:private-mem}
\end{equation}

Now, we represent the \textit{memory energy credit} of $\mathcal{A}$ on $s$ as the fraction of private memory of $\mathcal{A}$:
\begin{equation}
    \left(C^\text{DRAM}_\mathcal{A}\right)^s = \left[ \left( M_\mathcal{A} \right)^s \big/ \left( M_\text{total} \right)^s \right]^\sigma, \label{eq:mem-credit}
\end{equation}
where, similar to $\gamma$ in Eq.~\ref{eq:cpu-credit}, $\sigma$ is the machine-specific scaling factor.
With the formulated memory credit in Eq.~\ref{eq:mem-credit}, the DRAM energy attribution of $\mathcal{A}$ can be computed by:
\begin{equation}
    E_\mathcal{A}^\text{DRAM} = \sum_{s \in S} \left(E_\Delta^\text{DRAM}\right)^s \cdot \left(C^\text{DRAM}_\mathcal{A}\right)^s + \left(E_\text{static}^\text{DRAM}\right)^s. \label{eq:mem}
\end{equation}
\section{\systemname{}: Prototype Implementation} \label{sec:impl}
    % \begin{figure}[t]
    %     \centering
    %     \begin{adjustbox}{width=.8\linewidth,center=0pt}
    %       \includegraphics[width=\linewidth]{example-image}
    %     \end{adjustbox}
    %     \caption{Architecture of \systemname{}}
    %     \label{fig:arch}
    % \end{figure}
    
To evaluate our theoretical model, we develop and open source\footnote{\url{https://github.com/HongyuHe/energat}} a prototype implementation of our attribution model, named \systemname{}.
% It can be either imported as a client-side library by users or injected by cloud providers running alongside the user programs as a daemon (e.g., side-car containers). % and tracing their energy provenance.
% This design makes it suitable for energy-in-the-loop clouds (Fig.~\ref{fig:future}) in which energy attribution is exported from the end users for software optimization, billing, and energy-aware cluster operations.

% \subsection{Architecture of \systemname{}} \label{subsec:arch}

Firstly, to cleanly distinguish the energy consumption of the tool and the user program (\S\ref{sec:sw-acct}), we implement \systemname{} as a \textit{separate process} of the target application. Once it obtains the static power information, its 
main process creates a daemon \textit{thread} that samples the RAPL meter and relevant thread-level NUMA events every $T_\text{sample}$ time (Eq.~\ref{eq:ex-total}). 
\ed{Table~\ref{tab:counters} lists the sampled counters and their corresponding metrics used by the attribution model.
Apart from the maximum domain values and the clock speed that are obtained once at the beginning, all other counters are sampled at this frequency.}
This sampling period currently is set to 10 ms and can be adjusted based on the type of application it targets.
Note that, even the smallest sampling interval supported by RAPL (1 ms) is larger than the minimum scheduling granularity of the Linux kernel.
Nevertheless, \systemname{} only aims for an approximation of the conditional probabilities (Eq.~\ref{eq:residence-cpu} and \ref{eq:residence-mem}) with low measurement overheads. 
Given our evaluation results~(\S\ref{sec:eval}), the 10 ms interval appears empirically sufficient for precise energy attribution. 

Based on the statistics collected by the daemon thread, the parent process of \systemname{} computes the thread-level resource usages $\left( T_\mathcal{A}^\text{CPU} \right)^s \text{ and } (M_\mathcal{A})^s$ aggregated by sockets (Eq.~\ref{eq:cpu-time} and \ref{eq:private-mem}) .
It then calculates the CPU and memory energy credits by Eq.~\ref{eq:cpu-credit} and \ref{eq:mem-credit}.
Finally, \systemname{} attributes energy based on the energy credits (Eq.~\ref{eq:cpu} and \ref{eq:mem}) and stores energy traces in its database, which could be queried  later. 
\enlargethispage{\baselineskip}
% To facilitate adoption, we provide both a Python API and a command line interface.

\section{Attribution Model Evaluation} \label{sec:eval}
    
    % \begin{figure}[t]
    %     \centering
    %     \begin{adjustbox}{width=1.1\linewidth,center=0pt}
    %       \includegraphics[width=\linewidth]{figures/validation.pdf}
    %     \end{adjustbox}
    %     \caption{Validation by summation for our fine-grained energy attribution model. The model is indirectly validated if the light gray part plus the black portion is equal to the total value.}
    %     \label{fig:validation}
    % \end{figure}

    % \begin{figure}[t]
    %     \centering
    %     \begin{adjustbox}{width=1.1\linewidth,center=0pt}
    %       \includegraphics[width=\linewidth]{figures/cpucredit.pdf}
    %     \end{adjustbox}
    %     \caption{Change of energy credit ($C^\text{CPU}_\mathcal{A}$, Eq.~\ref{eq:cpu-credit}) assigned to the target application (left), reacting to the start of a ``neighbor task'' running the same workload as the target on the same host, while the actual amount of energy attributed to the target remains mostly unaffected (right).}
    %     \label{fig:credit}
    % \end{figure}

    % \begin{figure}[t]
    %     \centering
    %     \begin{adjustbox}{width=1.1\linewidth,center=0pt}
    %       \includegraphics[width=\linewidth]{figures/neighbor.pdf}
    %     \end{adjustbox}
    %     \caption{Comparison of total energies attributed to the target application with and without a noisy-neighbor task.}
    %     \label{fig:neighbor}
    % \end{figure}

This section presents preliminary evaluation results of our energy attribution model~(\S\ref{sec:method}) implemented in \systemname{}.
We employ the Linux benchmarking tool \texttt{stress-ng}~\cite{king2017stressng,fieni2020smartwatts} to create four types of workloads~(Table~\ref{tab:benchmarks}).
The aim of these microbenchmarks is to (1) cover different utilization levels of the two devices and (2) emulate the noisy-neighbor effect.
The testbed we use in the following experiments is a dual-socket server, where each NUMA node has 8 Intel Xeon E5-2630 CPUs (16 logical cores) and 32 GiB of DRAM.
We run each workload for one hour, averaging results over five runs.
% namely, \texttt{cpu} that focuses on energy used by compute, \texttt{mem} that exercises DRAM with various activities, and \texttt{mix} that is a mix of the first two types.

\begin{figure}[t]
        \centering
        \begin{adjustbox}{width=1.15\linewidth,center=0pt}
          \includegraphics[width=\linewidth]{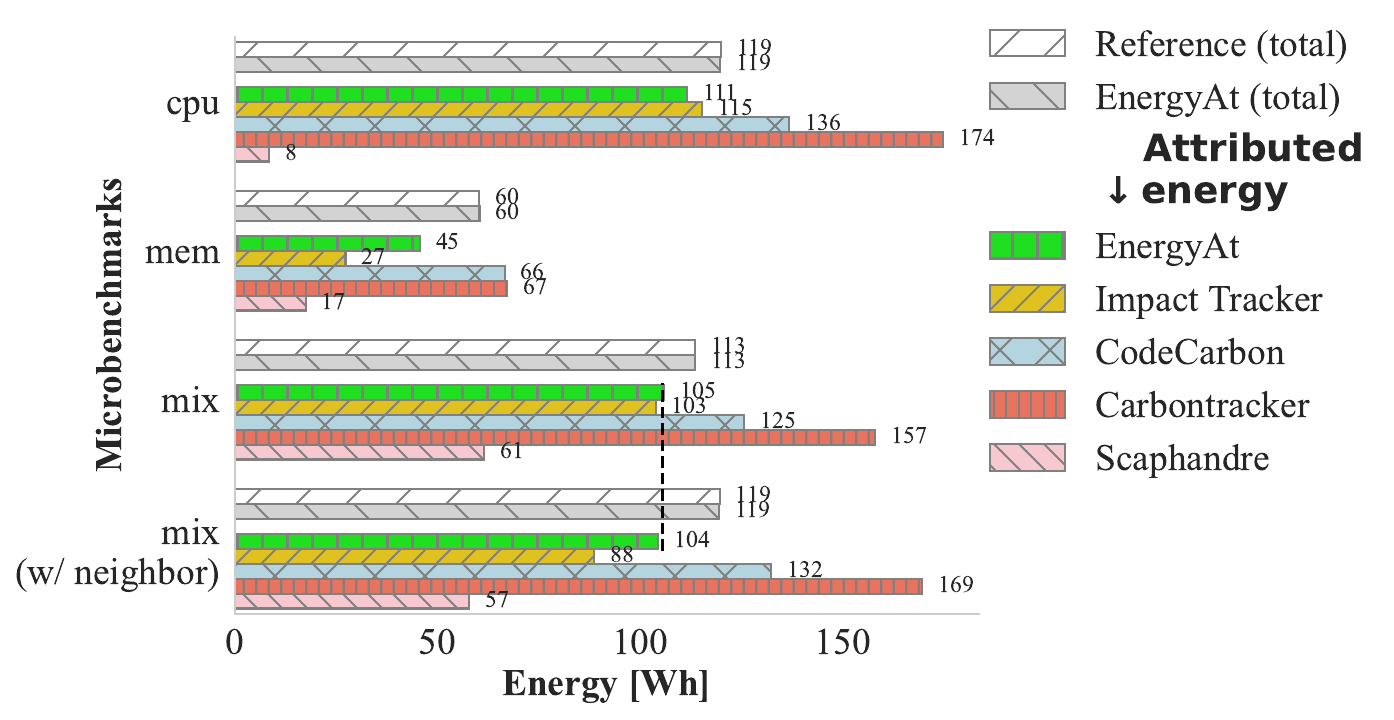}
        \end{adjustbox}
        \caption{Drastically different results from various tools on four microbenchmarks (Table~\ref{tab:benchmarks}). The bars below the \texttt{(total)} values are the attributed energies to the corresponding benchmark tasks by different tools. We highlight three observations: (1) existing methods can exhibit more than 46.3\% overestimation and 93.3\% underestimation of the \textit{attributed} energy compared to the \textit{total} values, (2) the total energy consumption measured by our fine-grained energy attribution model \systemname{} matches that of the reference value, and (3) the dotted line illustrates that \systemname{} is robust to the noisy-neighbor effect.}
        \label{fig:tools}
    \end{figure}

\vspace{-3pt}
\paragraph{\textbf{Model validation.}}
We start by validating the total energy consumed by the host server measured by \systemname{}. 
To this end, we make a popular Firefox plugin~\cite{firefox} run independently for reference.
As shown in Fig.~\ref{fig:tools}, the total energy consumption measured by \systemname{} closely matches that of the Firefox plugin. 
% The bars below the \texttt{(total)} values are the attributed energy to the corresponding benchmark tasks by different tools, which exhibit substantial over/under-estimations compared to the total values.

    \begin{table}[t]
    \centering
    \begin{adjustbox}{width=1.15\linewidth,center=0pt}
    \begin{tabular}{ l|l} 
    \toprule
    \textbf{Counters} & \textbf{Metrics} \\ 
    \midrule
    Intel RAPL package and DRAM domains & \makecell[l]{Accumulated energy consumption of CPU packages \\ and DRAM (through the \texttt{sysfs} interface)} \\
    \hline
    Intel RAPL maximum counter values & \makecell[l]{Maximum ranges of each domain for detecting \\ and mitigating counter overflow} \\
    \hline
    Memory statistics from the \texttt{numactl} package & \makecell[l]{Total, used, and private memory statistics for processes \\ and the operating system on a per-NUMA-node basis} \\
    \hline
    \texttt{/proc/*/task/*/stat} & \makecell[l]{User and kernel times for each task and its children} \\
    % \hline
    % \texttt{/sys/devices/system/cpu/*} & \makecell[l]{Cores to CPU socket mapping} \\
    \hline
    \texttt{CLK\_TCK} value & \makecell[l]{Number of clock ticks per second} \\
    \bottomrule
    \end{tabular}
    \end{adjustbox}
    \caption{Descriptions of metrics from sampled counters.}
    \label{tab:counters}
    \vspace{-15pt}
\end{table}
    \begin{table}[t]
    \centering
    \begin{adjustbox}{width=1.15\linewidth,center=0pt}
    \begin{tabular}{ c|l} 
    \toprule
    \textbf{Workload} & \textbf{Description} \\ 
    \midrule
    \texttt{cpu} & \makecell[l]{Sweeps CPU utilization from 0 to 100\% with equal \\ numbers of processes and threads loaded with matrix operations} \\
    \hline
    \texttt{mem} & \makecell[l]{Sweeps memory usage from 0 to 100\% with \\ one process continuously calling \texttt{mmap}/\texttt{unmap}} \\
    \hline
    \texttt{mix} & \makecell[l]{Keeps both the CPU and memory utilization at 50\% \\ using the same methods as that of \texttt{cpu} and \texttt{mem}} \\
    \hline
    \texttt{mix (w/ neighbor)} & \makecell[l]{Launches two \texttt{mix} workloads, treating one as the target application \\ and another as the ``noisy neighbor'' collocated on the same host} \\
    \bottomrule
    \end{tabular}
    \end{adjustbox}
    \caption{Descriptions of employed microbenchmarks.}
    \label{tab:benchmarks}
    \vspace{-15pt}
\end{table}
    
\begin{figure}[t]
        \centering
        \begin{adjustbox}{width=1.\linewidth,center=0pt}
          \includegraphics[width=\linewidth]{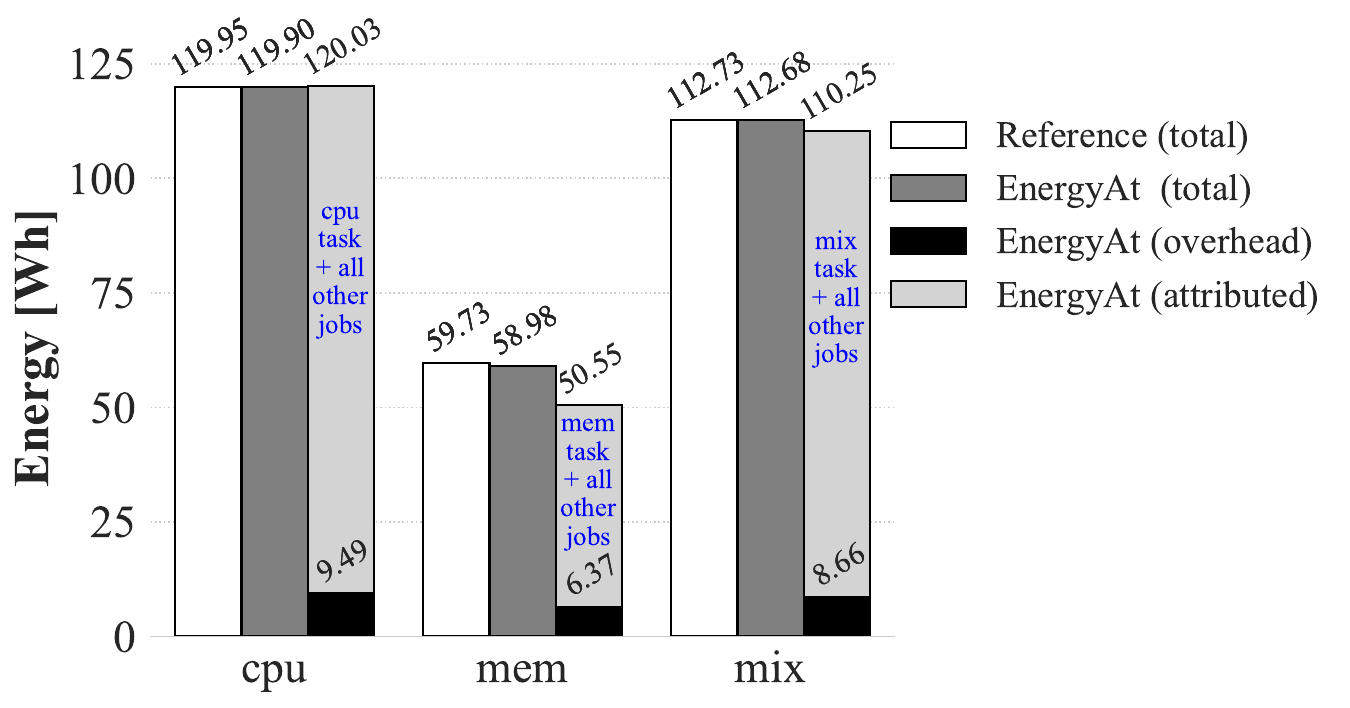}
        \end{adjustbox}
        \caption{Validation by summation for our fine-grained energy attribution model. The model is indirectly validated if the light gray part plus the black portion is equal to the total value (white or dark gray bars).}
        \label{fig:validation}
    \end{figure}

Next, we validate our fine-grained energy attribution model.
Since direct validation is infeasible~\cite{DBLP:conf/asplos/powercontainer}, we conduct validation by summation~\cite{DBLP:conf/asplos/powercontainer}.
In Fig.~\ref{fig:validation}, we trace the energy provenance of \textit{all} jobs present on the host, including the corresponding microbenchmarks, shown in light gray, and sum their attributed energies together (light gray bars) to compare with the total energy value of the machine (white and dark gray bars).
In other words, the attribution model is indirectly validated if the energy attribution of all the jobs, plus the energy used by \systemname{} (black bars), amounts to the same value as that of the total energy consumed by the host server.

This validation is non-trivial due to two main factors. 
Firstly, the power model is non-linear in nature. Secondly, each entity (thread/process) is traced independently. In other words, the attributed energies cannot simply be summed up to match the total energy consumption if the model fails to accurately assign energy to each entity individually.
% In practice, we let \systemname{} trace \texttt{systemd} as the target application, and it will automatically include all tasks that it has spawned and that it will spawn.
The resulting summation of attributed energies from all collocated tasks closely matches the total values on all three workloads (Table~\ref{tab:benchmarks}) with an average error margin of 4.52\%.
We anticipated a bit higher error for the \texttt{mem} workload, as \systemname{} currently only considers private memories in attribution and disregards any shared memories~(Eq.~\ref{eq:residence-mem}), which we defer to future work.

\begin{figure}[t]
        \centering
        \begin{adjustbox}{width=1.1\linewidth,center=0pt}
          \includegraphics[width=\linewidth]{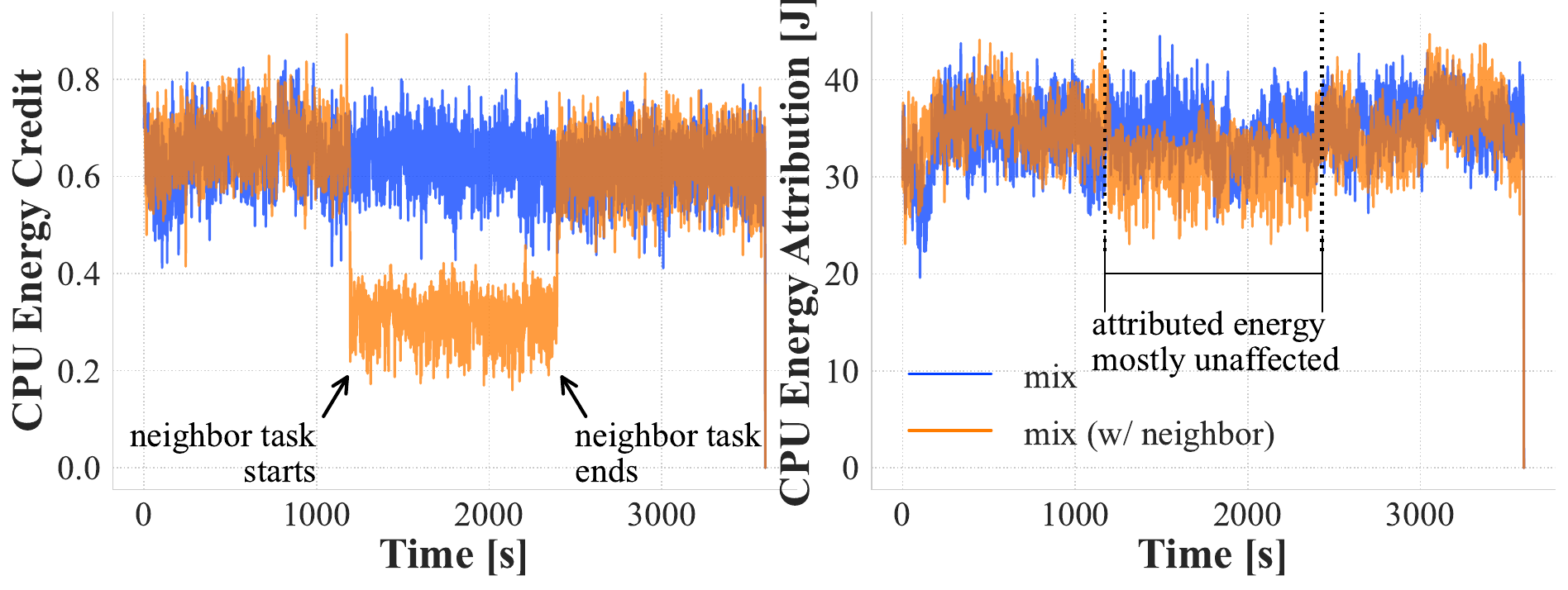}
        \end{adjustbox}
        \caption{Change of energy credit ($C^\text{CPU}_\mathcal{A}$, Eq.~\ref{eq:cpu-credit}) assigned to the target application (left), reacting to the launch of a ``neighbor task'' running the same workload as the target application on the same host, while the actual amount of energy attributed to the target remains mostly unaffected (right).}
        \label{fig:credit}
    \end{figure}

\vspace{-3pt}
\paragraph{\textbf{Robustness to noisy-neighbor effect.}}
Now, we evaluate the robustness and the effectiveness of the energy crediting method (Eq.~\ref{eq:cpu-credit} and \ref{eq:mem-credit}) in the presence of the noisy-neighbor effect.
To this end, we employ the \texttt{mix (w/ neighbor)} microbenchmark (Table~\ref{tab:benchmarks}).
Although energy consumption will be slightly different between runs even on the same machine with the same workload, we expect that the total energy attributed to the same workload should be approximately the same, regardless of whether it runs in an isolated environment or co-locate with a neighboring task.
When the neighbor task starts, the total energy consumption of the server increases due to higher resource usage, and the energy credit assigned to the target also drops, since the relative fraction of resources used by the target application reduces (Fig.~\ref{fig:credit} left).
This reactive change in energy credit assignment assures that the energy attributed to the target by \systemname{} remains almost unaffected (Fig.~\ref{fig:credit} right and Fig.~\ref{fig:tools}), while the measurements from other attribution tools change substantially due to the noisy-neighbor effect. 

\vspace{-3pt}
\paragraph{\textbf{Low attribution overhead.}}
We aim for not only a precise energy attribution method but also a low-overhead one for practical usage. 
% The lower the overhead, the more likely users and cloud providers are willing to adopt it.
Thus, unlike existing methods, we explicitly separate the energy used by \systemname{} from the traced application and its subtasks (\S\ref{sec:impl}).
When tracing the energy provenance of a single application (Fig.~\ref{fig:tools}), the energy overhead of \systemname{} is 6.5\% on average. 
The energy cost is 8.9\% on average when tracing all jobs on the server (Fig.~\ref{fig:validation}).
\enlargethispage{\baselineskip}
% In the worst case where it traces all the tasks on the whole server (Fig.~\ref{fig:validation}), the overhead is around 8.93\% on average.

\section{Limitations} \label{sec:limit}

Although our proposed method demonstrates promising results, several limitations are crucial to be addressed in future work.
First, the power modeling (Eg.~\ref{eq:cpu} and \ref{eq:mem}) does not take into account other relavent factors, such as various I/O and caches~\cite{DBLP:conf/asplos/powercontainer}.
Also, it only considers private memories, which results in lower energy attribution on the \texttt{mem} microbenchmark~(Fig.~\ref{fig:validation}).
This shortcoming is in part due to the restricted hardware interface and the overhead of sampling corresponding counters at a fine-grained level.
For instance, unlike well-integrated RAPL interfaces, accounting disk and network I/O requires external power meters. 
Similarly, obtaining \textit{both} per-thread and per-socket cache events is non-trivial.
Additionally, setting machine-specific model parameters ($\gamma,\delta$) currently needs hand-tuning for a certain platform.
Second, the energy cost of the prototype is non-negligible, which can be partially ascribed to the inefficiency of reading various counters.

Moreover, although \systemname{} automatically pins its process and daemon thread to the least-loaded core, it could still impose a higher performance penalty, compared to the coarse-grained tools.
While being an inherent tradeoff between granularity and cost, this drawback could be mitigated through a more optimized implementation of the proposed attribution model.
For example, the importance of various counter values differs by specific use cases~\cite{colmant2015process}, and in turn, they should be sampled at different granularity to reduce the overhead.
Lastly, validating fine-grained energy attribution model remains to be a prominent challenge. 
Validation by summation~\cite{DBLP:conf/asplos/powercontainer}~(\S\ref{sec:eval}) is rather limited in that it cannot provide insight into the energy attribution of each traced entity.
Since fine-grained validation is virtually impossible in a real system, full-system simulation (e.g., gem5~\cite{lowe2020gem5}) could be of help for this purpose.

\section{Challenges and Opportunities} \label{sec:future-work}

\begin{figure}[t]
        \centering
        \begin{adjustbox}{width=1.2\linewidth,center=0pt}
          \includegraphics[width=\linewidth]{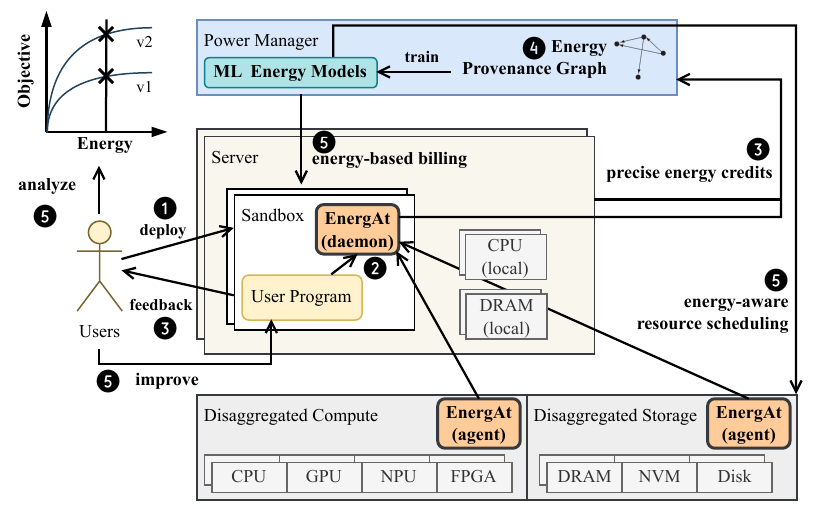}
        \end{adjustbox}
        \caption{Depiction of a sustainable cloud environment where both developers and providers make informed decisions based on \textit{distributed, fine-grained} energy attribution.}
        \label{fig:future}
    \end{figure}

We lay out a vision of sustainable cloud environments (Fig.~\ref{fig:future}) where \systemname{} can help both the users and cloud providers incorporate energy into their operation and development cycles.

In a sustainable cloud environment, service providers can inject \systemname{} into user sandboxes (e.g., as a side-car container), upon application deployment~(\circleblack{1}).
The tool runs alongside the user program as a daemon, monitoring the power dynamic by collecting performance counters and energy readings from the local machine and from remote agents located on disaggregated compute and/or storage nodes~(\circleblack{2}).
Then, \systemname{} attributes energy to the user application and reports corresponding energy credits to the Power Manager of the cluster, informing the application owner of the energy consumption~(\circleblack{3}).
The precise energy credits can be used to construct provenance graphs~\cite{pasquier2017provenance1,pasquier2018provenance2,kruit2020tab2know}~(\circleblack{4}) for tracing the power relationships among deployed applications. 
Such graph representations can be leveraged to train high-quality ML models that facilitate power management.
Furthermore, users can analyze and improve their software energy efficiency based on the detailed feedback~(\circleblack{5}).
Similarly, cloud providers can make energy-aware decisions accordingly~(\circleblack{5}), e.g., energy-based billing and workload migration for mitigating hot spots.
That said, there are as many promising opportunities as there are challenges in this virtuous cycle.
% By injecting \systemname{} into user sandboxes (e.g., as a side-car container) for instance, cloud providers do not need to treat user jobs as black boxes and can obtain detailed energy profiles.
% Different from coarse-grained power statistics collected from, e.g., the Intelligent Platform Management Interface (IPMI), fine-grained energy attribution would enable cloud providers to construct provenance graphs~\cite{pasquier2017provenance1,pasquier2018provenance2} for energy consumption and make energy-aware decisions, e.g., migrating workloads with known power dynamics to mitigate hotspots.
% That said, there exist as many challenges as there are promising opportunities.
% users can be informed of their application behaviors related to power consumption and, in turn, allowing them to optimize for energy efficiency iteratively.
% Similarly, instead of collecting centralized, coarse-grained energy statistics, cloud providers can use the fine-grained information gathered from end users,

\vspace{-3pt}
\paragraph{\textbf{New interfaces for secure and efficient energy reporting.}}
% \TODO{security}
The lack of secure and efficient interfaces between hardware and software severely inhibits energy measurement~(\circleblack{2}).
For instance, reading RAPL requires manual timestamp alignment~\cite{DBLP:journals/sigmetrics/shortcodepath} and privileged permission for security~\cite{leak}.
Moreover, virtualization is a similar challenge as energy-related counters are generally not propagated to the virtualized applications.
Consequently, most of the energy models for virtualized environments are predictive and treat user programs as black boxes~\cite{fieni2020smartwatts,bohra2010vmeter,krishnan2011vmmetering,dhiman2010predictvirtual}.
\ed{Thus, tracing fine-grained energy provenance amid layers of virtualization is way more complex and remains an open question.}

% \vspace{-3pt}
% \paragraph{More fine-grained energy attribution.}
% \TODO{
% Decouple shared infra cost;
% Thread- \textit{and} socket-level cache events
% }

\vspace{-3pt}
\paragraph{\textbf{Energy attribution for new cloud computing services.}}

Multi-tenancy and heterogeneity in the cloud are being taken to the extreme in order to remain profitable in the post-Moore's Law era.
For example, various hardware accelerators are increasingly shared among large numbers of tenants.
They have fundamentally different inner workings compared to Von Neumann architectures.
Furthermore, high-level cloud services (e.g., DBaaS, MLaaS, and FaaS) have emerged and are popularized.
Their abstractions are farther away from hardware and highly distributed in nature.
These factors not only pose challenges to collecting energy statistics~(\circleblack{2}) but also to tracing precise energy provenance~(\circleblack{3}, \circleblack{4}).
\enlargethispage{\baselineskip}

% Coupled with the trend of compute-storage disaggregation, these factors pose challenges to software energy attribution. 
% For example, the distributed nature of these new paradigms significantly complicates the matter~\cite{seo2009distributedtrace,anand2022odd}, hardware accelerators have fundamentally different inner workings compared to traditional Von Neumann architectures, and, service-level energy tracing (e.g., for FaaS, MLaaS, DBaaS) remains an open question due to their high levels of abstraction.

% \vspace{-3pt}
% \paragraph{Tracing energy provenance amid layers of virtualization.}
% Virtualization is foundational to cloud computing.
% It, however, induces substantial challenges in energy attribution.
% For instance, the ability to access energy-related registers/counters in virtualized environments depends on whether the underlying platform (e.g., hypervisors) traps and propagates such information to the hosted applications.
% As of the current writing, few public clouds provide such capabilities.
% Consequently, most energy models for virtualized environments treat containers and VMs as black boxes and predict their power dynamics based on kernel events~\cite{fieni2020smartwatts,bohra2010vmeter,krishnan2011vmmetering,dhiman2010predictvirtual}.

\vspace{-3pt}
\paragraph{\textbf{NUMA-aware energy optimization.}}

\ed{
Although this work shows the importance of carefully attributing energy for applications running on multiple sockets, how developers and cloud providers can leverage the proposed the model~(\circleblack{5}) remains unexplored.
For instance, is the energy consumption of an application the same whether it runs on a different core of the same socket or on a different socket within the same server?
The answer to this question would be useful for both users and cloud providers in terms of improving energy efficiency.
This question, however, is also challenging as there are temperature effects that impact the actual power usage.
}

\vspace{-3pt}
\paragraph{\textbf{Improving software performance with energy in mind.}}

With the availability of new abstractions and tools, developers can gain insights into the energy dynamics of their applications~(\circleblack{3}).
In turn, development decisions should not only focus on traditional performance optimization but also consider energy efficiency~(\circleblack{5}).
For instance, a 10\% increase in throughput may not be considered beneficial if it comes at the cost of a 50\% higher power consumption.
Similarly, in the context of training ML models, energy should be taken into account as an early-stopping criterion, since a mere 0.1\% loss reduction may not justify the addition of 100 kWh of energy consumption.
\ed{Last but not least, it is worth revisiting the energy efficiency of traditional algorithms and data structures whose optimization has primarily focused on performance.
% Taking databases as an example, the preference between a B-tree or hash table could be different when not only considering their performance but also energy consumption on a certain hardware platform.
}
% \enlargethispage{\baselineskip}

% \TODO{Add performance overhead measurement}

% \begin{itemize}
%     % \item Energy measurements require privileged access
%     % \item Energy propagation in layers upon layers of virtualization
%     % \item GPU, FPGA tracing and attribution
%     \item Performance cost
%     \item New metrics
%     \item Benchmark DB workloads
% \end{itemize}

% \newpage
\begin{acks}
We thank the anonymous reviewers for their many helpful comments. We especially would like to thank Shail Dave for his invaluable assistance throughout the revision process. Shail's extensive feedback on each section and meticulous attention to detail, including the revision of every figure, played a vital role in enhancing the quality of the final paper.
\end{acks}

\balance
% \newpage
\bibliographystyle{ACM-Reference-Format}
\bibliography{reference}

\appendix
\end{document}